\documentclass[aps,prd,reprint,twocolumn,superscriptaddress,longbibliography,nofootinbib,floatfix,showpacs]{revtex4-1}
\usepackage{epsfig}
\usepackage{amsmath,amssymb,amsfonts}
\usepackage{hyperref}
\usepackage{mathrsfs}
\usepackage{bbm}
\usepackage{slashed}
\usepackage{graphicx}
\usepackage{verbatim}
\usepackage[usenames]{color}

\newcommand{\be}{\begin{equation}}
\newcommand{\ee}{\end{equation}}
\newcommand{\bi}{\begin{itemize}}
\newcommand{\ei}{\end{itemize}}
\newcommand{\bea}{\begin{eqnarray}}
\newcommand{\eea}{\end{eqnarray}}

\newcommand{\bra}[1]{\langle\,#1\,|}          

\newcommand{\ud}{\mathrm{d}}

\newcommand{\LCm}{{\scriptscriptstyle -}} 
\newcommand{\LCp}{{\scriptscriptstyle +}}
\newcommand{\LCpm}{{\scriptscriptstyle \pm}}

\newcommand{\LCperp}{{\scriptscriptstyle \perp}}

\newcommand{\Eins}{\mathbbmss{1}}

\newcommand{\A}{C}

\newcommand{\tdf}{{\widetilde{f'}}}
\newcommand{\bw}{\begin{widetext}}
\newcommand{\ew}{\end{widetext}}

\begin{document}

\title{Infra-red divergences in plane wave backgrounds}
\author{Victor Dinu}
\email[]{dinu@barutu.fizica.unibuc.ro}
\affiliation{Department of Physics, University of Bucharest, P.O.\ box MG-11, M\v{a}gurele, 077125 Romania}

\author{Thomas Heinzl}
\email[]{theinzl@plymouth.ac.uk}
\affiliation{School of Computing and Mathematics, University of Plymouth PL4 8AA, UK}

\author{Anton Ilderton}
\email[]{anton.ilderton@physics.umu.se}
\affiliation{Department of Physics, Ume\aa\ University, SE-901 87 Ume\aa, Sweden}

\begin{abstract}
We show that the emission of soft photons via nonlinear Compton scattering in a pulsed plane wave (laser field) is in general infra-red divergent. We give examples of both soft and soft-collinear divergences, and we pay particular attention to the case of crossed fields in both classical and quantum theories.

\end{abstract}
\pacs{}
\maketitle

\section{Introduction}
There is growing interest in the use of strong laser fields to probe both physics beyond the standard model \cite{Redondo:2010dp, Jaeckel:2010ni}, and the high-intensity, low-energy regime of QED \cite{610731,DiPiazza:2011tq,Heinzl:2008an}. Much of this interest is focussed on quantum effects such as non-perturbative pair creation from the vacuum \cite{Dunne:2008kc} and vacuum birefringence \cite{Heinzl:2006xc}. As the centre-of-mass energies in laser-particle scattering are typically low compared to accelerator experiments (laser frequencies are in the range $1$-$10^4$ eV  \cite{ELI,XFEL}) quantum effects are hard to detect. However, the high intensity, or flux, of the laser can compensate to some extent for the low energy \cite{Heinzl:2009zd}.

The high field strengths of modern lasers require a nonperturbative approach, with most progress having been made when the laser fields are described by a plane wave (this model holds for beams which are not too tightly focussed \cite{Heinzl:2009nd,Harvey:2012ie}). In this case, scattering  amplitudes can be calculated for arbitrary field strengths using Volkov propagators and wavefunctions \cite{Volkov:1935,Reiss1,Nikishov:1963,Nikishov:1964a,Narozhnyi:1964}. Using this model, increasingly complex semi-classical (tree level) processes are being studied; examples include Compton scattering within the background field \cite{Hartin}, M\o ller scattering \cite{Elk}, trident pair production \cite{Ilderton:2010wr} and the multi-photon emission processes
\be\label{MULTI-PHOTON}
	e^-(p)\xrightarrow{\text{in laser}} e^-(p') + \gamma(k^1) + \gamma(k^2) + \ldots \gamma(k^n) \;.
\ee
The $n=1$ process is well-studied \cite{Harvey:2009ry,Boca:2009zz,Seipt:2010ya, Mackenroth:2010jr,Hartin:2011vr}. It goes by the name `nonlinear Compton scattering' since, for low background field strengths, the scattering amplitude becomes a sum over ordinary Compton amplitudes for each frequency in the background, see below. The case $n=2$ has been considered in \cite{Lotstedt:2009zz, Seipt:2012tn}.  The low field strength amplitude for this process is proportional to that of two-photon Compton scattering in ordinary QED,
\be
	e^-(p) + \gamma(k_\text{laser})\rightarrow e^-(p') + \gamma(k^1) + \gamma(k^2) \;,
\ee
which is infra-red (IR) divergent \cite{Jauch}. The divergence is inherited by the full $n=2$ process.

In \cite{DiPiazza:2010mv}, an incoherent-sum approximation was used to study large $n$, and it was observed that the (tree level) probabilities $\mathbb{P}(n)$ for the processes (\ref{MULTI-PHOTON}) can exceed unity. Investigating this statement is not entirely straightforward since, while it is conceptually trivial to calculate $S$-matrix elements for arbitrary $n$, it is computationally exhausting and multiple numerical integrations are required to obtain integrated probabilities\footnote{See \cite{Titov:2012rd} for recent progress in stimulated pair production, obtained from nonlinear Compton by crossing symmetry}. Some intuition can however be obtained by studying the classical limit, in which the probabilities $\mathbb{P}(n)$ can be calculated exactly; one finds \cite{Boca:2009zz,Linda},
\be\label{LIMIT}
	\lim_{\hbar\to 0} \mathbb{P}(n) = \frac{1}{n!}(N_\gamma)^n \;,
\ee
where $N_\gamma$ is the classically obtained `number of photons' emitted by a particle passing through a plane wave (see below). There is no {\it a priori} reason why $N_\gamma$ should be smaller than one. For the $n=0$ process, `scattering without emission', the total (tree level) probability is
\be\label{why}
	\mathbb{P}(0) \equiv 1 \; ,
\ee
without approximation and for arbitrary initial electron momentum and an arbitrary plane wave. (This can only be correct in the absence of the background field.) Hence, the total probability of photon emission via (\ref{MULTI-PHOTON}) exceeds unity. Something is clearly wrong, but it is not hard to identify the origin of the problem and the higher order corrections which will resolve it; (\ref{LIMIT}) is the archetypal relation associated to the infra-red problem. Let us therefore recall how IR divergences arise, and are dealt with, in QED. (For a particularly lucid introduction to the IR see \cite[\S 6]{Peskin:1995ev}.)

One first encounters the IR problem in the process of bremsstrahlung, which is the emission of a photon from an electron as it passes a nucleus. The emitted photon can be arbitrarily soft and this leads to an IR divergence, already at tree level, when the emitted photon frequency goes to zero. However, any detector has a finite resolution and therefore cannot distinguish between sufficiently soft bremsstrahlung and scattering without emission; the Bloch-Nordsieck result is that, when one considers the physically measurable sum of the probabilities of these two IR divergent processes, the IR divergence cancels between them \cite{Bloch}. IR divergences are not solely the domain of bremsstrahlung, though. Ordinary Compton scattering is infrared divergent at one-loop level. To cancel this divergence one must account for both ordinary Compton scattering and double-Compton scattering of one hard and one soft photon \cite{Brown:1952eu}. In general, then, physically measurable quantities which account for detector resolutions and experimentally indistinguishable processes are IR finite \cite{Bloch,Kinoshita:1962ur,Lee}.

The IR divergences in the $n=2$ case of (\ref{MULTI-PHOTON}) have been dealt with to date by inserting a cutoff or damping factor. Nor does it seem to have been noticed that nonlinear Compton scattering, $n=1$, can be IR divergent depending on the properties of the chosen background field: as we will show, a divergence arises when the plane wave's physical fields (not its potential) contain a Fourier zero mode. Such pulses are called `unipolar' \cite{produce} and, physically, the presence of the zero mode means the pulse can transfer a {\it net acceleration} to a particle. At this point we should recall the `Lawson-Woodard theorem' which states (loosely) that vacuum solutions of Maxwell's equations cannot transfer net energy to a particle,  {\it provided} the $v\times B$ force vanishes, the interaction time is infinite, there is no radiation reaction, and that the particle is always ultra-relativistic \cite{woodward:1946,woodward:1948,palmer:1987}. We deal here with pulsed plane waves, which do exhibit the $v\times B$ force, have a finite interaction time, and can be arbitrarily shaped, giving particle arbitrary and time dependent velocities. Thus, Lawson-Woodard does not apply \cite{Hartemann-reply}.

Vacuum acceleration in a real laser field is not only possible (see \cite{Salamin:2002gh} for high-order Gaussian beam calculations and \cite{Esarey:1995cw} for acceleration methods using multiple beams) but has been observed \cite{observed}. Since plane waves provide the most accessible and best understood models of laser fields, we should clearly allow for plane waves which model accelerating structure. The IR divergences we discuss will also appear in more realistic geometries: it is a good idea to understand the simplest model first. Furthermore, we note that the problematic result (\ref{why}) holds even for `ordinary'  (non unipolar) plane waves.

The purpose of this paper is therefore to carefully examine the origin and severity of IR divergences in plane wave backgrounds, starting with the classical theory and then considering tree level quantum calculations. The presentation is intended to be pedagogical. In a sequel paper, the divergences will be removed by calculating the appropriate higher order corrections.

The paper is organised as follows. We begin in Sect.~\ref{SEC:CLASSICAL} with a general discussion of classical radiation and the `IR catastrophe'. In Sect.~\ref{SEC:PLANE} we apply this to the particular case of a plane wave background and show how the IR sector of the emission spectrum is related to the net energy transferred by the plane wave. In Sect.~\ref{SEC:VOLKOV}, we turn to QED. LSZ reduction formulae are derived for unipolar plane waves, and this produces Volkov solutions with the correct boundary conditions. Using these, nonlinear Compton scattering is addressed in Sect.~\ref{SEC:NLC} and shown to be IR divergent. We investigate the soft and perturbative limits, and compare with both bremsstrahlung and ordinary Compton scattering.  We also discuss the seemingly contradictory example of crossed fields (constant plane waves), which can accelerate but do not lead to an IR divergence. We will see that existing literature results are indeed IR finite, but that they describe an unphysical scattering process which has little to do with experiment. We conclude in Sect.~\ref{SEC:CONCS}.

\section{\label{SEC:CLASSICAL} Classical radiation: IR behaviour}
%
%
Consider a particle moving in an arbitrary background field. All that is required for a covariant evaluation of the particle radiation is its trajectory $x^\mu$ as a function of proper time $\tau$. From this trajectory, one forms the Fourier transformed current,
\be \label{JMU}
  j^\mu (k') = e \int\! \ud\tau \, u^\mu (\tau) \, e^{ik' \cdot x(\tau)} \; ,
\ee
where $u^\mu(\tau) \equiv \dot{x}^\mu(\tau) $ is the four-velocity of the particle. Next, one uses  Poynting's theorem and the Lienard-Wiechert potential \cite{Mitter:1998}, or the advanced/retarded Green's  functions \cite{Coleman:1961zz}, to obtain the following expression for the four-momentum $P^\mu$ of the radiation field:
\be
	P^\mu = -\frac{1}{2}\int\!\frac{\ud^4 k'}{(2\pi)^3} \, \text{sign} (k^{\prime 0}) \, \delta (k^{\prime 2}) \, k^{\prime \mu} \, j (k')\!\cdot\! j^*(k') \; .
\ee
Integrating over $k^{\prime 0}$ and employing the on-shell delta function, the energy $P^0$ may be written
\be\label{ENERGY}
  P^0 = \int\! \ud\omega' d\Omega\ \omega' \rho(k') \; ,
\ee
where we have introduced the frequency $\omega'$ and the spectral density  $\rho$ measuring the `number of photons' radiated per unit frequency per unit solid angle,
\be \label{RHO}
  \rho(k') = - \frac{\omega'}{2(2\pi)^3} \, j(k') \cdot j^* (k') \; .
\ee
We now take a closer look at the physics of these expressions. This will also be good preparation for the quantum calculation.

\subsection{Current conservation}
A little care must be taken in defining the Fourier transformed current $j_\mu(k')$. The na\'ive definition  (\ref{JMU}) does not obey current conservation,
\be
   k'\!\cdot\! j(k') = -ie \int\!\ud\tau \, \frac{\ud}{\ud\tau} \, e^{ik' \cdot x(\tau)} \not=0 \;.
\ee
since the boundary terms do not in general cancel each other. To understand why, go back to (\ref{JMU}) and assume that the background field turns on and off at finite times. We parameterise the path such that the particle enters the field at proper time $\tau=0$ and position $x^\mu=0$, with momentum $p_\mu$. (We can shift these initial conditions arbitrarily, and the pulse can turn on/off arbitrarily smoothly: our final result will be completely general.) The particle then exits the pulse at some proper time $\tau=\tau_f$, at some position $x^\mu_f$ and with some momentum $p'_\mu$, all determined by the classical equations of motion. With this, the current becomes
\bea
	\nonumber  j^\mu (k') &= \displaystyle\frac{e}{m}p_\mu \int\limits_{-\infty}^0\! \ud\tau \, e^{ik'.p \tau/m} +e\int\limits_0^{\tau_f}\! \ud\tau \, u^\mu (\tau) \, e^{ik' \cdot x(\tau)} \;  \\
	   &+\displaystyle \frac{e}{m}p'_\mu \int\limits^{\infty}_{\tau_f}\! \ud\tau \, e^{ik'.[x_f + p' (\tau-\tau_f)/m]} \;.
\eea
The first and third terms do not behave well in the IR, i.e.\ at large distances, where the phases diverge\footnote{The divergent phases we will encounter in this paper are, as we will see, related to soft divergences and not the similarly named `phase divergences' which occur in, say, pair creation \cite{Bagan:1999jf,Bagan:1999jk}.} The integrals can be regulated using an $i\epsilon$ prescription in the exponents, giving
\be\label{this}
\begin{split}
	 j^\mu (k') =  ie\, \frac{e^{ik'.x_f}\, p'_\mu}{k'.p'+i\epsilon} &-ie\, \frac{p_\mu}{k'.p-i\epsilon} \\
	&+  e\int\limits_0^{\tau_f}\! \ud\tau \, u^\mu (\tau) \, e^{ik' \cdot x(\tau)} \;.
\end{split}
\ee
The first and second terms now give the (boosted) Coulomb fields of the particle before and after interaction with the background, as follows from inserting (\ref{this}) into (\ref{ENERGY}) and carrying out the $k'_0$ integral, see \cite[\S 6]{Peskin:1995ev}. Letting $\epsilon\to 0$, the current (\ref{this}) is easily checked to be conserved, $k'.j(k')=0$. Integrating (\ref{this}) by parts, the boundary terms cancel the Coulomb terms and we obtain our final, compact result
\be\label{this2}
\begin{split}
	 j^\mu (k') &= -e\int\limits\! \ud\tau \, e^{ik' \cdot x(\tau)} \frac{\ud}{\ud\tau} \bigg(\frac{u^\mu(\tau)}{i k'.u} \bigg) \;,
\end{split}
\ee
where the integral is {\it automatically} restricted to the pulse duration since the integrand goes like the acceleration $\dot u$, as is seen by expanding the derivative,
\be\label{JMU2}
  j^\mu (k') = -e \int\!\ud\tau \, e^{ik' \cdot x(\tau)} \, \frac{\dot{u}^\mu u^\nu - \dot{u}^\nu u^\mu}{i (k' \cdot u)^2} \, k^{\prime\nu} \;.
\ee
This shows manifestly that only accelerated charges radiate, see \cite[\S 14]{Jackson}. In other words, calculating the spectral density (\ref{RHO}) using (\ref{JMU2}) corresponds to measuring only the radiation caused by the action of the external field, and not the intrinsic (boosted) Coulomb fields of the particle from before and after scattering. This relates the spectral density to the quantity of interest in the quantum theory, which is the spectrum of photons emitted in the interaction of an electron with the background. Note that since $u_\mu(\tau)$ is timelike and $k'_\mu$ is lightlike, the denominators in (\ref{this})--(\ref{JMU2}) are nonzero unless $\omega' = 0$; we address this now.

\subsection{The classical IR problem}
Curing the large distance problems in the expression (\ref{JMU}) has lead us to (\ref{this2}), or (\ref{JMU2}), which makes physical sense and makes the `classical IR catastrophe' manifest. This problem is often stated in terms of the `total number of emitted photons', defined by
\be\label{IR-zero}
	N_\gamma := \int\! \ud\omega' d\Omega\ \rho(k') \;.
\ee
The IR problem is that $N_\gamma$ diverges at low frequencies:
\be\label{IR-one}
	N_\gamma \sim \int\limits_0 \!\ud \omega'\ \frac{\text{const.}}{\omega'} \;.
\ee
Since the `number of photons' is not a classical concept, we rephrase (\ref{IR-one}) in terms of energy. An equivalent statement is that the energy (\ref{ENERGY}) emitted at low frequency {\it is independent of frequency}, i.e.\
\be \label{IR-two}
	P^0 \sim \int\limits_0\! \ud\omega'\ \text{const.}
\ee
There is therefore no divergence in any measurable classical object, but it is the behaviour (\ref{IR-two}) which signals a corresponding IR divergence in the quantum theory. Let us now make (\ref{IR-zero}) to (\ref{IR-two}) concrete. Consider the emission of low frequency radiation. We write $k'_\mu = \omega' n'_\mu$, and expand (\ref{this2}) for small $\omega'$. We find
\be\begin{split}\label{SOFT-J}
	 j^\mu (k') &= \frac{ie}{{\omega'}} \int\limits_0^{\tau_f}\! \ud\tau \, \frac{\ud}{\ud\tau} \bigg(\frac{u^\mu(\tau)}{n'.u} \bigg)  +\mathcal{O}(\omega'^0) \\
	 	&= \frac{ie}{{\omega'}}  \bigg(\frac{{p'}^\mu}{n'.p'}-\frac{p^\mu}{n'.p} \bigg) + \mathcal{O}(\omega'^0) \;.
\end{split}
\ee
(The same result follows immediately from (\ref{this}), since the integral term is bounded and can be dropped in the soft limit.) Again, $p'_\mu$ ($p_\mu$) is the particle's momentum when it leaves (enters) the pulse. The spectral density therefore has the low-frequency expansion
\be
	\rho(k') =  -\frac{1}{2(2\pi)^3 \omega'}  \bigg(\frac{{p'}}{n'.p'}-\frac{p}{n'.p} \bigg)^2 + \mathcal{O}(\omega'^0) \;,
\ee
from which the behaviour in (\ref{IR-one}) and (\ref{IR-two}) follows,
\be\begin{split} \label{IR-CLASSICAL}
	N_\gamma &= -\frac{\alpha}{(2\pi)^2} \bigg[\int\!\ud\Omega \bigg(\frac{{p'}}{n'.p'}-\frac{p}{n'.p} \bigg)^2\bigg] \int\limits_0 \!\ud \omega'\ \frac{1}{\omega'} \;, \\ 
	P^0 &= -\frac{\alpha}{(2\pi)^2} \bigg[\int\!\ud\Omega \bigg(\frac{{p'}}{n'.p'}-\frac{p}{n'.p} \bigg)^2\bigg] \int\limits_0\! \ud\omega'\ 1 \;, 
\end{split}
\ee
to lowest order and where $\alpha=e^2/4\pi$. The denominators here are {\it strictly} positive. The angular integral can be performed exactly and is non-zero. In the limit that the background field provides a `sudden kick', instantaneously changing the particle's momentum, the expressions (\ref{IR-CLASSICAL}) become exact. There is no sudden kick here, in general, as we have said nothing about the properties of the background field, which is the statement that ``the precise form of the trajectory \ldots does not affect the low-frequency radiation" \cite[\S 6]{Peskin:1995ev}. We now apply these general results to the case of plane waves.

\section{Plane waves} \label{SEC:PLANE}
%
A plane wave travelling in the negative $z$-direction is characterised by the lightlike vector $n_\mu=(1,0,0,1)$ and some scale $\omega$ which is usually (but not necessarily) the dominant frequency of the wave. We write $k_\mu:=\omega n_\mu$. The transverse electric fields $E_j$ ($j=1,2$) depend arbitrarily on the dimensionless, Lorentz invariant variable $\phi:=k.x$, which can be identified with lightfront time. Lightfront variables are defined via $\phi = k.x = k_\LCp x^\LCp$ where $x^\LCpm = t \pm z$, $x_\LCpm = (x_0 \pm x_3)/2$ and $x^\LCperp=\{x^1,x^2\}$. The field strength may be written
\be\label{PW:FIELD}
	F_{\mu\nu}(k.x) = f'_j(k.x) (k_\mu a^j_\nu-a^j_\mu k_\nu)  \;,
\ee
where the $f'_j$ are profile functions describing the shape of the electromagnetic fields and, for our choice of $k_\mu$, the polarisation vectors become $a^1_\mu=(a_0m/e)(0,1,0,0)$ and $a^2_\mu=(a_0m/e)(0,0,1,0)$:  we normalise the profile functions $f'_j$ such that $(f_j'f_j')_\text{rms}=1$, sum over $j$, rms taken over the whole pulse, so that the parameter $a_0$ is always equal to $a_0 \equiv e E_\text{rms} / m\omega$ \cite{Heinzl:2008rh}. The energy in a pulse of duration $T$ is then proportional to $a_0^2 T$.

A particle in a plane wave, neglecting radiation reaction, has kinetic momentum $\pi_\mu \equiv m u_\mu$ obeying the Lorentz force equation
\be \label{EOM}
  \dot{\pi}_\mu = \frac{e}{m}F_{\mu\nu}(k.x) \pi^\nu \;.
\ee
It follows that $k.\pi$ is conserved and one can trade proper time for lightfront time $\phi$. In complete generality, we assume that the particle is free, with momentum $p_\mu$, until some lightfront time $\phi_i$ when it first encounters the field. The corresponding solution of the Lorentz equation is
\be \label{UMU}
  \pi_\mu (p ;\phi) := p_\mu - e {\A}_\mu(\phi) + \frac{2 e {\A}(\phi)\!\cdot\!p - e^2 {\A}^2(\phi)}{2 k\!\cdot\! p} \, k_\mu \; .
\ee
Here, ${\A}_\mu$ is the integral of the field strength,
\be\label{ADEF}
	{\A}_\mu(\phi) := a^j_\mu \int\limits_{\phi_i}^\phi\!\ f'_j(\varphi) =:  a_\mu^j f_j(\phi) \;.
\ee
It is easy to check both that (\ref{UMU}) obeys the correct initial condition, $\pi_\mu(p;\phi_i) = p_\mu$, and that $\pi^2\equiv m^2$. When the pulse turns off at, say, $\phi=\phi_f$ the particle again becomes free. By definition, the function $C_\mu$ then becomes constant, i.e.,
\be
	C_\mu(\phi_f)=C_\mu(\infty),
\ee
which we write as $C_\mu^\infty$ from here on. Note that $C_\mu^\infty$ is a vector of {\it Fourier zero modes} of the electromagnetic field strengths, i.e.\
\be
	\A_\mu^\infty \equiv a^j_\mu \tilde{f'_j}(0) \;.
\ee
Fields for which the Fourier zero mode is non-vanishing are called unipolar, one example of which is a subcycle pulse. Unipolar pulses can be produced from `ordinary' fields with a vanishing zero mode by interaction with a nonlinear optical medium, see \cite{produce}.

\subsection{Plane waves and the infra-red}
For our purposes, all plane wave fields fall into one of two categories, defined by whether the electromagnetic field's Fourier zero mode is zero or not:
\be\label{FINT}
	\int\limits_{\phi_i}^{\phi_f}\!\ud \phi\, F_{\mu\nu}(\phi)  \begin{cases} = 0  \iff \A^\infty_\mu= 0\,, & \text{`whole-cycle',} \\
												\not=0 \iff \A^\infty_\mu \not=0\,, & \text{`unipolar'.}
										\end{cases}
\ee
A particle entering a `whole-cycle' field with momentum $p_\mu$ leaves with the same momentum, i.e.\ experiences no net acceleration, since ${\A^\infty_\mu}=0$ in (\ref{UMU}) and therefore
\be
	\pi_\mu(p;\phi) = p_\mu \quad \text{when} \quad \phi\geq \phi_f \;.
\ee
These are pulses which, in a loose sense, contain a `whole' number of cycles. The same particle entering a unipolar field is accelerated, leaving with a different momentum $\pi_\mu(p;\infty)$,
\be\label{difference}
	\pi_\mu(p;\infty) = p_\mu - e {{\A}}_\mu^\infty + \frac{2e{\A}^\infty.p-e^2{\A}^\infty\!\cdot\!{\A}^\infty}{2k.p} k_\mu \;,
\ee
which differs from $p_\mu$ in both the transverse ($\perp$) and lightfront energy (lower  $+$) components. This is the precise form of the `Lawson-Woodward theorem' for plane waves. These results hold independently of both the pulse duration and details of its field structure; we are not discussing unphysical edge effects.

$C^\infty_\mu$ will play a crucial role in what follows. It neatly encodes a property of the {\it field strength}: mathematically, the Fourier zero mode of the electromagnetic fields and, physically, the ability of the electromagnetic fields to do net work on a particle\footnote{We note that the zero mode can also be obtained from the gauge invariant phase of a lightlike Wilson loop \cite{Liu:2006he}. We note further that acceleration, through `sudden kicks', relates IR divergences to cusp singularities in Wilson loops, see \cite{Bagan:1999jk} and references therein}.

\begin{figure}[t!]
\includegraphics[width=0.95\columnwidth]{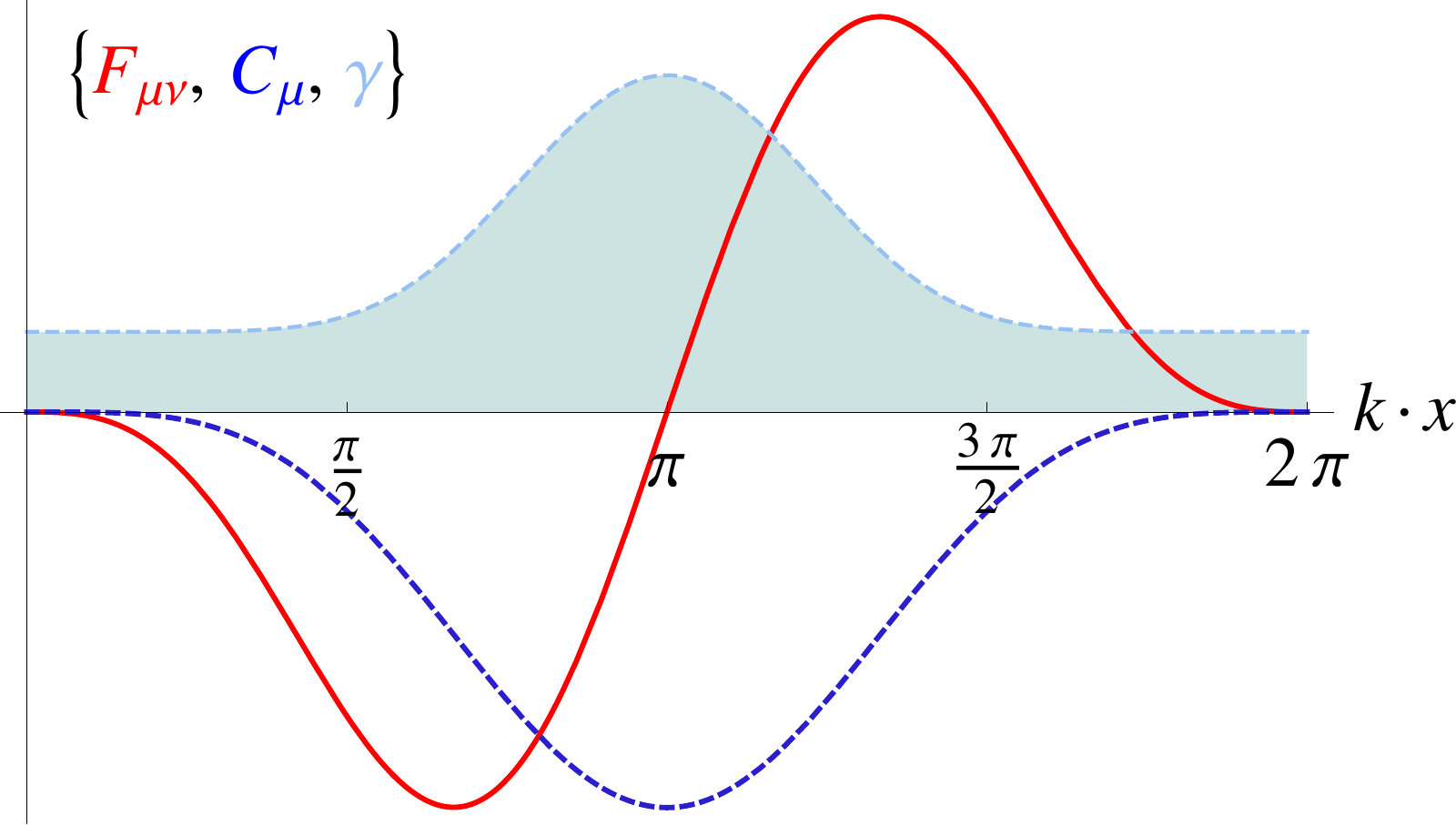}
\includegraphics[width=0.95\columnwidth]{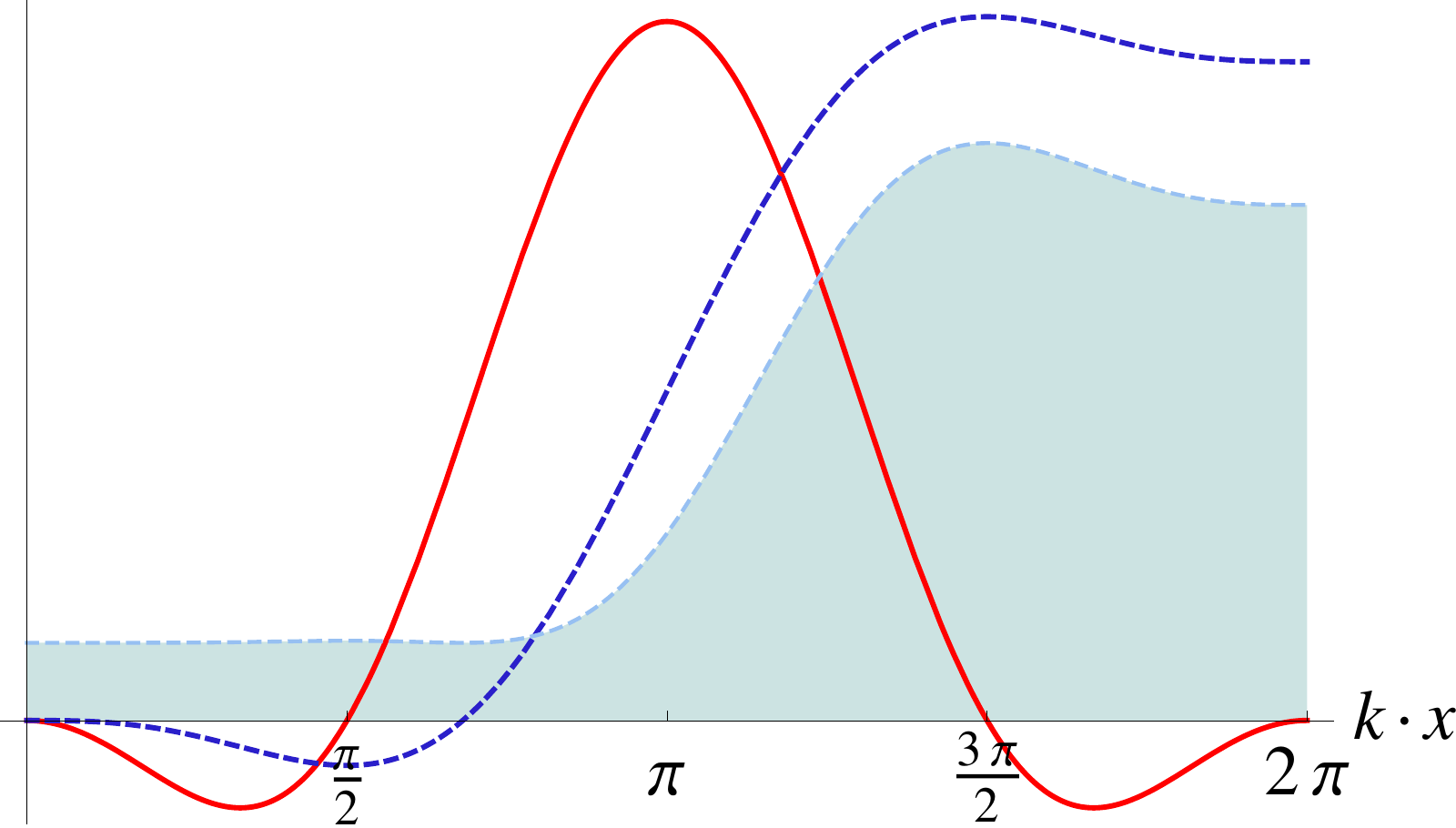}
\caption{\label{EXEMPEL} {\it Not to scale.} The $\gamma$-factor [filled] of an electron in a plane wave with field strength $F_{\mu\nu}$ [red/solid] and potential $C_\mu$ [blue/dashed]. {\it Upper panel:} In a whole-cycle pulse, the $\gamma$-factor returns to its initial value when the electron leaves the field. {\it Lower panel:} In a unipolar pulse, the electron gains a net acceleration, signalled by a non-zero potential at the end of the pulse. This potential term yields the non-zero boundary term of (\ref{SOFT-J}) which signals the soft IR divergence.}
\end{figure}

An electron passing through a `whole-cycle' pulse acquires no net acceleration, so incoming and outgoing momenta are equal, $p' = \pi(p;\infty) = p$. In particular,
\be
	\frac{{p'^\mu}}{n'.p'}-\frac{p^\mu}{n'.p} \equiv 0 \;,
\ee
and so the leading order terms in (\ref{IR-CLASSICAL}) vanish: the classical number of photons $N_\gamma$ becomes IR finite and the low energy spectrum is frequency {\it dependent}.  The implication is that the corresponding quantum processes are IR finite, and this is born out: nonlinear Compton scattering contains no IR divergence provided the pulse contains a whole number of cycles, see \cite{Boca:2009zz,Seipt:2010ya,Mackenroth:2010jr} for examples. The typical situation for whole-cycle pulses is sketched in Fig.~\ref{EXEMPEL}, top panel.

Consider now an electron passing through a unipolar pulse. The electron leaves this pulse with a net acceleration, $p' = \pi(p;\infty) \not= p$, because of the non-vanishing Fourier zero mode $C^\infty_\mu$.  The typical situation is sketched in Fig.~\ref{EXEMPEL}, lower panel; the electric field will clearly push the particle more in one direction than the other, giving a net acceleration. A simple way to model such pulse shapes shape is to employ a carrier phase, see the appendix for details. Since $C^\infty\not=0$, the boundary term of (\ref{SOFT-J}) is non-zero, and this gives a divergent photon number in (\ref{IR-CLASSICAL}). We therefore expect nonlinear Compton scattering to exhibit the usual IR divergence of QED when the background field has unipolar structure. We confirm this below. We note that even an {\it infinitesimal} deviation from whole-cycle structure in the field strength is enough to cause an IR divergence, so it is really unipolar rather than whole-cycle pulses which are the general case. There is also a special case, which we consider before turning to the quantum theory.

\subsection{Soft and collinear divergences}
`Crossed fields' are constant and homogeneous electric and magnetic fields of equal magnitude and orthogonal polarisation, in other words, constant plane waves. They provide one of the most common models of laser fields outside of monochromatic beams; the probabilities of nonlinear Compton scattering and stimulated pair production in crossed fields form the basis of cascade codes, for example \cite{Elkina:2010up,Bulanov:2010gb}.

\begin{figure}[t!]
\includegraphics[width=\columnwidth]{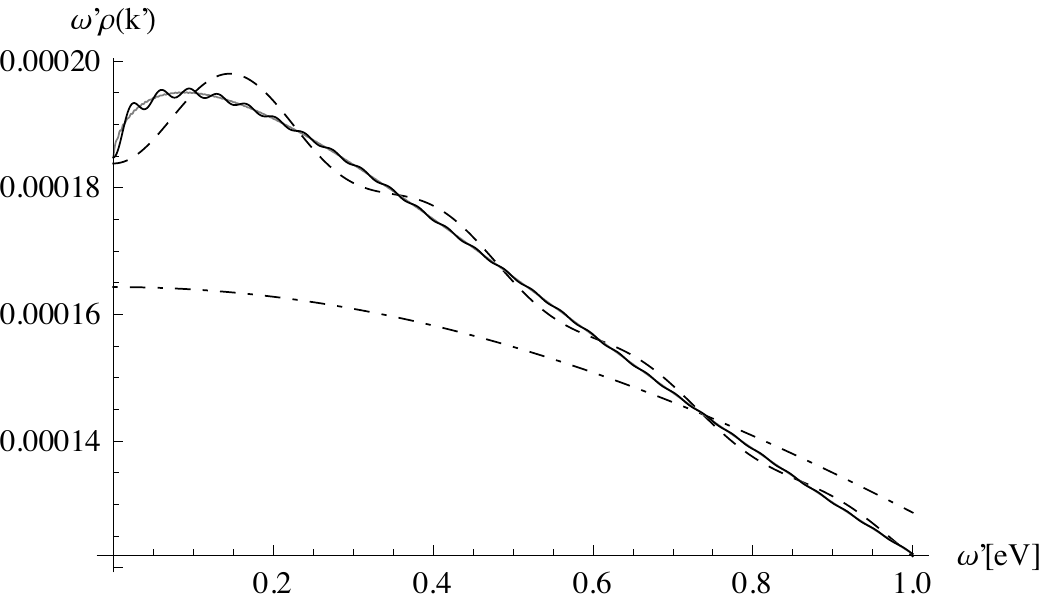}
\includegraphics[width=0.95\columnwidth]{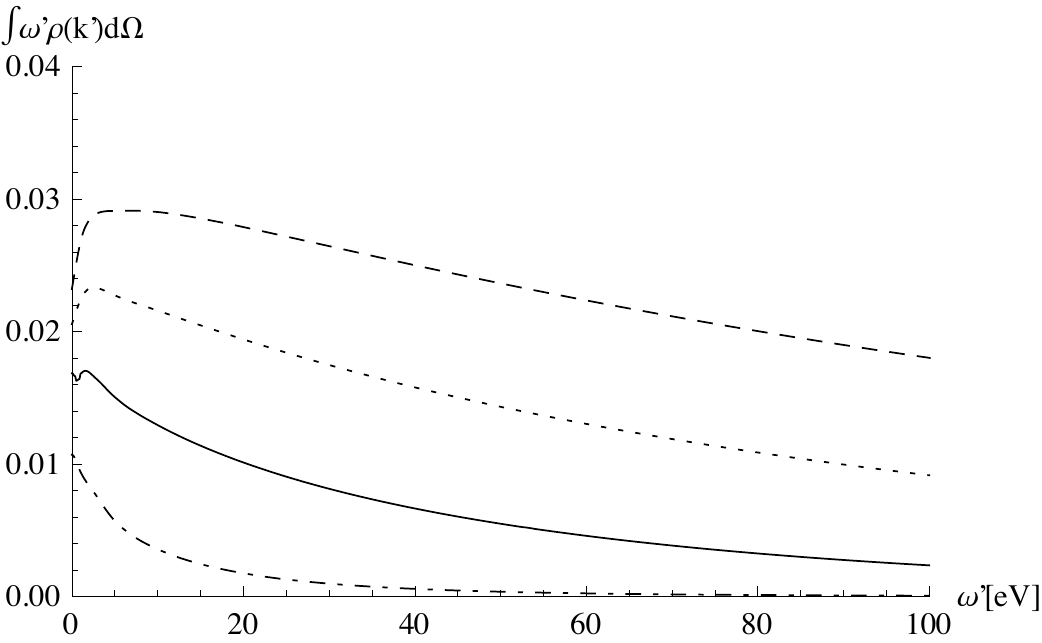}
\caption{\label{CROSSED-PLOTS} The energy spectrum for a crossed field (with $E/E_S = 2.10^{-6}$). {\it Upper panel:} Fixed emission angles, $\theta=\varphi=\pi/2$, for duration $T=2,5,10,20$ (dot-dashed, solid, dotted, dashed). At $\omega'=0$, the spectrum converges to the result implied by (\ref{SOFT-J2}) as $T$ increases. {\it Lower panel:} duration $T=5,10,15,20$. When integrated over emission angles, the soft limit includes the developing soft-collinear divergence in (\ref{SOFT-J2}) and increases like $\log T$ as $T\to\infty$. }
\end{figure}

The infinite extent of crossed fields is somewhat unphysical. To study their infrared properties in a controlled manner we therefore consider a plane wave which is constant for $-\tfrac{T}{2}<\phi<\tfrac{T}{2}$ and otherwise zero. The definition (\ref{ADEF}) then gives
\be\label{CROSSED-POTENTIAL}
	\A_\mu = a^1_\mu\begin{cases}
				0 & \phi < -\tfrac{T}{2} \\
				\phi+T/2 & -\tfrac{T}{2} \leq \phi < \tfrac{T}{2} \\
				T & \phi \geq \tfrac{T}{2} \;.
			\end{cases}
\ee
Clearly this field accelerates, since $C_\mu^\infty = T a_\mu^1$, which implies a log-divergent photon number. If we focus on the soft sector, evaluating (\ref{SOFT-J}) in the limit that $T\to\infty$ yields
\be\begin{split}\label{SOFT-J2}
	 j^\mu (k') &=  -ie\bigg(\frac{{k}^\mu}{k'.k}-\frac{p^\mu}{k'.p} \bigg) + \mathcal{O}(\omega'^0) \;,
\end{split}
\ee
which is independent of the chosen field strength $E$.  As well as the soft divergence, we also have here a `soft and collinear' divergence when $k'_\mu \propto k_\mu$. Collinear divergences are known to appear only in association with massless particles (for their removal see \cite{Kinoshita:1962ur,Lee,Lavelle:2005bt}). The reason they can appear here is that any constant electric field, when allowed to persist for an infinite time, accelerates all incoming particles to the speed of light. In this sense, the final state particles are effectively `massless' (as in high energy approximations, for example, in which one neglects mass terms compared to momentum terms). Indeed, the dominant term in the particle's final momentum for large $T$ is,
\be\label{SOFT-J3}
	\pi_\mu(p;\infty) \sim T^2 k_\mu + \mathcal{O}(T) \;,
\ee
which is lightlike, and the replacement of $\pi_\mu(p;\infty)$ with $k_\mu$ is manifest in (\ref{SOFT-J2}). These results are summarised in Fig.~\ref{CROSSED-PLOTS}, where we plot the energy density $\omega'\rho(\omega')$. At fixed emission angles, the value at $\omega'=0$ is non-zero, illustrating the soft divergence, and converges to (\ref{SOFT-J2}) as the duration increases. When the emission angle is integrated out, the low frequency value grows with $T$ because of the developing collinear divergence. The growth rate can be found analytically for $\omega'=0$, where the angular integrals can be performed exactly. One finds,
\be
	\int\!\ud\Omega\ \omega' \rho(\omega') \big|_{\omega'=0}= \frac{e^2}{\pi^2}\log T  + \ldots,
\ee
which is logarithmic. So, crossed fields lead to both soft and soft-collinear divergences in the photon number: their IR structure is {\it worse} than the generic case. Surprisingly, the literature results for the quantum case, ie.\ for nonlinear Compton scattering in crossed fields, are IR {\it finite}. This rather stark contradiction will be resolved in Sect.~\ref{SEC:NLC}.

%
\section{Asymptotic states and Volkov solutions}\label{SEC:VOLKOV}
%
%
Consider QED coupled to an additional, external gauge field $\A_\mu$. (The doubling of notation is deliberate, but for now $\A_\mu$ is arbitrary.) We briefly outline how one can calculate in the theory when the background is treated nonperturbatively. The action is
\bw
\be \label{SPLIT}
	S = \int\!\ud^4x \ -\frac{1}{4}F^{\mu\nu}F_{\mu\nu} + \bar\psi \big[i \gamma^\mu(\partial_\mu + ie \A_\mu)-m\big] \psi\ + \text{ gauge fixing + sources } \bigg|_{\hspace{-11.5pt}\leftarrow\ \rightarrow} -e\bar\psi \slashed{A} \psi \;,
\ee
\ew
where $A_\mu$ is the dynamical (quantised) photon field and $F_{\mu\nu}$ is its field strength. Everything to the left of the bar is considered to be `free', and everything to the right is `interacting'. With regards to the quantum fields, this is the same split as is made in perturbation theory, except that the `free theory' now contains a background field. Hence, the position space Feynman rules are unchanged from QED except that the fermion propagator is the inverse of $i\slashed{\partial}-e\slashed{C}-m$ rather than $i\slashed{\partial}-m$. If this propagator can be calculated exactly, the background field will be accounted for nonperturbatively. In order to convert Feynman diagrams into $S$-matrix elements one appeals to the usual LSZ reduction formulae: amputation converts external leg propagators into wavefunctions which are, for the fermion fields, solutions of the Dirac equation in the background field $C_\mu$.

It is only advantageous to use the split (\ref{SPLIT}) if the `free theory' can be solved exactly (i.e.\ if the propagator can be calculated exactly),  for otherwise one must in any case resort to perturbative methods. Plane wave backgrounds provide an example of such a theory: the propagator is known exactly for arbitrary plane waves \cite{Brown:1964zzb}, and amputation produces the well-known Volkov wavefunctions which are indeed solutions of the Dirac equation in a plane wave background, see below. These statements hold in (at least) the gauge for which we choose the gauge potential $C_\mu$ to be equal to the classical $C_\mu(k.x)$ we encountered in (\ref{ADEF}). This has the benefit of making the physics manifest. It is also what is, often implicitly, done in the literature, and is what we will do here. (Other gauges are available. The reader is invited to find one which can reproduce all known results in a simpler, faster way.)

We now point out a peculiarity. The Volkov solutions depend, aside from the usual exponential $p.x$ factors, {\it only} on the lightfront variable $\phi=k.x$. $S$-matrix elements therefore exhibit overall momentum conservation in the $x^\LCperp$ and $x^\LCm$ directions. This implies that the total $p_\LCperp$ and $p_\LCm$ of the incoming particles is conserved. In particular, there is no dependence anywhere on, say, $\A.x$, and hence it seems impossible for any $S$-matrix element to recover the transverse push proportional to $e\A_\infty$ exhibited in the classical theory, see (\ref{UMU}). To illustrate, the $S$-matrix element for scattering in a plane-wave without emission will have support on the conservation law
\be
	p'_\mu = p_\mu + s k_\mu \;,
\ee
where $s$ arises as the Fourier transform of the $\phi$ dependence introduced by the background. For all momenta being on-shell, this equation has only one solution, $s=0$. This is not the correct result for unipolar pulses, where we expect $p_\mu$ to become $\pi_\mu(p;\infty)$ (classically). We will show below that the resolution of this problem requires a careful, but straightforward, analysis of the large distance behaviour of our theory: it is therefore not surprising that this should be taken up before discussing the IR problem.

\subsection{LSZ reduction and Kibble's basis}
The usual LSZ assumption about the large distance behaviour of QED is that the interaction between the {\it quantised} fields switches off. It is well known that this assumption is responsible for IR divergences, so we expect them to persist here \cite{Kulish:1970ut,Horan:1999ba}. We do {\it not} assume that the coupling to the background field switches off, though. Indeed, there is a sense in which it does not: a particle can be accelerated by a unipolar field and, in the absence of other interactions, retains this `information', even though the pulse itself has switched off.

To account for this, we will consider what happens when the gauge potential becomes a {\it constant} in the far future, as is the case for unipolar fields, $\A_\mu \to \A_\mu ^\infty$. Our asymptotic theory (in the future) therefore consists of fermions minimally coupled to a constant gauge field $\A^\infty$.  This theory is free, since constant gauge fields are pure gauge. We write $D^\infty_\mu \equiv \partial_\mu + i e \A_\mu^\infty$. The electron solutions of the Dirac equation in such a background are
\be
	e^{-i(p'+e\A^\infty).x}u_{p'} \;,
\ee
where $p'_\mu$ obeys $p'^2=m^2$ and is the eigenvalue of $iD_\mu^\infty$: it is the kinematic momentum. It is now straightforward to go through the usual steps leading to the LSZ reduction formulae. Amputation for incoming particles is unchanged, since both the classical and quantum gauge fields switch off in the past. For outgoing electrons, though, LSZ reduction gives the following amputation instruction:
\be\label{AMP-OUT}
	-i\int\!\ud^4 x\ e^{i(p'+e\A_\infty).x}\, \bar{u}_{p'} \big( i\slashed{D}^\infty-m\big)_x \bra{0} T \psi(x)\ldots \;,
\ee
which differs from the usual result only in the presence of $C^\infty_\mu$. Applying (\ref{AMP-OUT}) to the Volkov propagator, one obtains the following expressions for the appropriate incoming and outgoing wavefunctions:
\begin{widetext}
\be\begin{split}\label{volkovs}
	\text{$e^-$ in:}\quad &{\Psi}^\text{in}_{p,\sigma}(x) := \bigg[\Eins + \frac{e}{2k.p}\, \slashed{k} \slashed{\A}(k.x)\bigg]u_{p}^\sigma\ \exp\bigg[-ip.x - \frac{i}{2k.p}\int\limits^{k.x}_{-\infty} 2e \A.p-e^2 \A^2\bigg]\;,\\
	\text{$e^-$ out:}\quad &{\bar\Psi}^\text{out}_{p',\sigma}(x) := \bar{u}_{p'}^\sigma\bigg[\Eins + \frac{e}{2k.p'} \, \delta \slashed{\A}(k.x) \slashed{k}\bigg] \exp\bigg[ i(p'+e\A_\infty).x - \frac{i}{2k.p'} \int\limits^{\infty}_{k.x} 2e\, \delta \A.p'-e^2 \delta \A^2\bigg] \;,
\end{split}
\ee
%
%
%
where $\delta \A_\mu(k.x) := \A_\mu(k.x) - \A^\mu(\infty)$. The limits on the integrals are not assumed, but follow as part of LSZ. So, incoming electrons are described by ordinary Volkov solutions while outgoing electrons are described by the second wavefunction in (\ref{volkovs}); it is straightforward to check that both satisfy the Dirac equation in the background $\A_\mu(k.x)$. Positrons solutions are obtained by sending $u\to v$ and $e\to-e$. Complemented with the usual propagator, the use of (\ref{volkovs}) completes the Feynman rules for the theory.

The corresponding wavefunctions for scalar particles have appeared in \cite{49713}. They were suggested as an alternative basis for outgoing states which would remove infinite phase factors from $S$-matrix elements. This does not quite work, though: while $\bar\Psi_\text{out}$ ($\Psi_\text{in}$) behaves well in the far future (past), it does not behave well in the far past (future), and the $S$-matrix element contains an integral over {\it all} times. Rather, the use of (\ref{volkovs}) makes the correct physics manifest, and the divergent phases are only removed by regulating the $S$-matrix elements themselves, as we will see.

Let us briefly check that these `new' LSZ rules describe the correct large distance behaviour of the theory. We return to the process of scattering without emission. Using (\ref{volkovs}), and going again to Fourier space by trading $k.x$ for dimensionless $s$, the $S$-matrix element for this process now takes the form (for some $F$ which we do not need explicitly)
\be\label{SFI-ELASTIC0}
	S_\text{no emission} = \int\!\ud s\ \delta^4(p'+e\A_\infty-p- s k) F\;.
\ee
There is again only one point of support for the delta function, as one finds by squaring the conservation law:
\be\label{ELASTIC-S}
	p'^2 = (p - e \A_\infty + s k)^2 \implies s = \frac{2e\A_\infty.p-e^2\A^2_\infty}{2k.p}\;.
\ee
Inserting this into (\ref{SFI-ELASTIC0}) we see that the $S$-matrix element for scattering without emission has support when $	{p'}^\mu = \pi^\mu(p;\infty)$, where we recognise the asymptotic kinematic momentum $\pi^\mu$ from (\ref{difference}).  In other words, the scattering amplitude now tells us that an electron experiences both the longitudinal and transverse pushes implied by the Lorentz force as it passes through a plane wave, as it should. This resolves the puzzle introduced in Sect.~\ref{SEC:VOLKOV} regarding the transverse terms in the momenta. Our LSZ analysis therefore yields the correct physics. We can now construct the $S$-matrix element for nonlinear Compton scattering, regulate it, and examine its IR structure.

\section{Nonlinear Compton scattering: IR divergence} \label{SEC:NLC}
\subsection{$S$-matrix element: regularisation}
Nonlinear Compton scattering, $e^-(p) \xrightarrow{\text{ in laser }} e^-(p') + \gamma(k')$, has the following $S$-matrix element to lowest order in the interaction between quantised fields (i.e.\ to tree level, with the background accounted for to all orders),
\be\label{SFI2}
	\begin{split}
		S_{fi} &= -ie\displaystyle\int\! \ud^4x\ {\bar\Psi}^\text{out}_{p',\sigma'}(x) \slashed{\varepsilon}e^{ik'.x} {\Psi}^\text{in}_{p,\sigma}(x) \\
		&= -\frac{ie}{2k_\LCp}(2\pi)^3\delta^3_{\LCperp,\LCm}(p' + e \A_\infty + k' - p)     \int\! \ud\phi \  e^{i\Phi(s_\LCp, \phi) }\ \text{Spin}(\phi) \;.
		\end{split}\raisebox{-30pt}{\includegraphics[width=0.15\columnwidth]{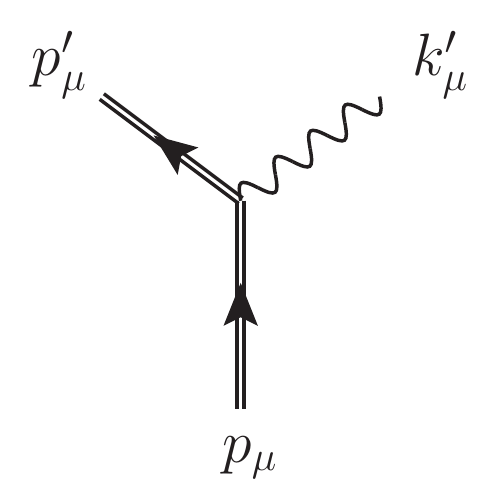}}
\ee
The $\Psi$'s are as in (\ref{volkovs}). To reach the second line, the integrals over $x^\LCperp$ and $x^\LCm$ are performed to yield delta-functions. The remaining integral over $x^\LCp$ is written as an integral over $\phi$.  `Spin' contains the photon polarisation and all the spin structure coming from the Volkov solutions (\ref{volkovs}), while $\Phi$ contains all the $\phi$-dependent phases coming from the same. These, together with $s_\LCp$, are given explicitly below. The $\phi$-integral in (\ref{SFI2}) needs to be regulated. Proceeding just as in the classical theory, we split the integral into three parts corresponding to before, during and after the pulse. Regulating with a damping factor essentially `cuts out' (in a gauge invariant way, as we confirm shortly) the before-and-after pieces of the $S$-matrix element in which no scattering can occur. The resulting expression is
\be
	\label{SFI3}
	S_{fi} = \frac{ie}{2k_\LCp}(2\pi)^3\delta^3_{\LCperp,\LCm}(p' + e \A_\infty + k' - p)     \int\! \ud\phi \  e^{i\Phi(s_\LCp, \phi) }\ \frac{\ud}{\ud\phi}\bigg[\frac{\text{Spin}(\phi)}{i\Phi'(s_\LCp,\phi)} \bigg]\;.
\ee
The dash on $\Phi$ is a derivative w.r.t.\ $\phi$. It is conceptually clearer to again Fourier transform, trading $\phi$ for a dimensionless variable $s$ which represents the lightfront energy taken from the background. The derivative of the term in square brackets is proportional to the background field strength, and hence the integrand vanishes outside the pulse. This means firstly that the Fourier transform is well defined and secondly that there are no infinite phase factors to worry about, as promised. The $S$-matrix element becomes
\be
	\label{SFI4}
	S_{fi} = ie\int\!\frac{\ud s}{2\pi}\ (2\pi)^4\delta^4(p' + e \A_\infty + k' - p-sk)\Gamma(s)  \;, \quad \text{with}\quad \Gamma(s):= \int\ud\phi\ e^{i\Phi(s,\phi)}\ \frac{\ud}{\ud\phi}\bigg[\frac{\text{Spin}(\phi)}{i\Phi'(s,\phi)} \bigg]\;.
\ee
Explicitly, the spin and phase parts are
\be
	\Phi(s,\phi) = s\phi - \int\limits_\phi^{\phi_f} \frac{2e\delta\A.p'-e^2\delta \A^2}{2k.p'} - \int\limits_{\phi_i}^\phi \frac{2e\A.p-e^2\A^2}{2k.p} \;,\qquad \text{Spin}(\phi)\ = {\bar u}^{\sigma'}_{p'}\bigg(1+\frac{e \delta \slashed{\A }\slashed{k}}{2k.p'}\bigg)\slashed{\varepsilon}\bigg(1+\frac{e\slashed{k}\slashed{\A}}{2k.p}\bigg)u^\sigma_p \;.
\ee
(Undoing the Fourier transform sets $s$ to a particular value $s_+$, but working in Fourier space allows us to maintain covariance, and the resulting expressions are clearer.) Before proceeding to the emission probability itself we should check that our regularisation is gauge invariant with respect to transformations of the quantum fields. This can be confirmed by showing that  (\ref{SFI4})  vanishes when $\varepsilon \to \varepsilon + \xi k'$: one finds that the resulting change in $\Gamma(s)$ is
\be\label{DELTA-GAMMA}
	\delta \Gamma(s) = \xi\, \bar{u}_{p'}\slashed{k}u_p \, \int\ud\phi\ e^{i\Phi(s,\phi)}\ \frac{\ud}{\ud\phi}\bigg[\frac{i\Phi'(s,\phi)}{i\Phi'(s,\phi)} \bigg] = 0\;,
\ee
as required. We can now wrap the incoming state into a wavepacket, normalised per unit lightfront volume (so the incoming particle carries a normalisation of $1/\sqrt{2p_\LCm}$ rather than $1/\sqrt{2p_0}$), square up the $S$-matrix element and obtain the total probability of emitting a photon, averaged over initial spins, summed over final spins and polarisations, as
\be\label{THE-PROB}
	\mathbb{P} = \frac{e^2}{2k.p} \int\!\ud f\ \int\!\frac{\ud s}{2\pi}\ (2\pi)^4\delta^4(p' + e \A_\infty + k' - p-sk)\ \frac{1}{2}\sum\limits_{\sigma,\sigma',\varepsilon}|\Gamma(s)|^2  \;.
\ee
As usual, the wavepacket drops out of the final expression, and the integral over final states is
\be
	\ud f = \frac{\ud^3p'}{(2\pi)^32p'_0}\ \frac{\ud^3 k'}{(2\pi)^32k'_0} \;.
\ee
Of the seven integrals in $\mathbb{P}$, four can be performed using the delta functions. Methods for evaluating the remaining three integrals are discussed in \cite{Seipt:2011dx}, see also \cite{Titov:2012rd}.
\end{widetext}
\subsection{Probability of emission: IR divergence}
We can now investigate the IR contribution to the probability (\ref{THE-PROB}).  Using the kinematics implied by the delta function in (\ref{SFI4}) or (\ref{THE-PROB}) one finds that the phase $\Phi$ has a single stationary point, corresponding to the point of soft emission, $\omega'=0$. The function $\Gamma(s)$ therefore diverges at this point. The classical analogue of this statement was that $k'.u\not=0$ unless $\omega'=0$; see (\ref{this2}) and the discussion following.  At the point of soft emission, the argument of the delta function in (\ref{THE-PROB}) becomes
\be
	p' + e\A_\infty -p - s k \to  0 \;,
\ee
which is just the inelastic scattering condition we found in (\ref{SFI-ELASTIC0}). In order to study the IR limit we therefore expand around (\ref{ELASTIC-S}), writing

\be
	s = t + \frac{2e\A_\infty.p-e^2\A^2_\infty}{2k.p} \;,
\ee
and look at the limit of small $t$. We eliminate the $p'$ integrals in (\ref{THE-PROB}) using the delta-functions. The remaining calculation is straightforward; the denominator $\Phi'$ becomes, for example,
\be
	\Phi'(\phi) = t + \frac{1}{k.p}[k'.\pi^\mu(p;\phi) - k'.\pi(p;\infty)\big] +\ldots
\ee
and in the the soft limit one has $\omega'_t = t {k.p}/{n'.\pi(p;\infty)} + \ldots$ and so $\omega'\propto t$. Hence, the remaining delta-function in (\ref{THE-PROB}) may be used to perform either the $\omega'$ integral or, equivalently, the $t$ integral. After performing the spin sums, the soft-photon contribution to the probability becomes
\be\label{IR-RESULT}
	\mathbb{P} = -\frac{\alpha}{(2\pi)^2}\int\!{\ud \Omega'} \bigg(\frac{\pi(p;\infty)}{n'.\pi(p;\infty)}-\frac{p}{n'.p}\bigg)^2\int\limits_{0}\!{\ud t}\ \frac{1}{t}  + \ldots
\ee
This diverges when $t\to 0$, i.e.\ at the point of soft emission. The probability is therefore IR divergent whenever the term in large brackets is non-zero, i.e.\ whenever the field is able to accelerate the particle, such that $\pi(p;\infty) \not= p$, which requires a unipolar field with nonvanishing $\A_\mu^\infty$. The singularity is logarithmic, as in bremsstrahlung, and depends only on whether $\A_\mu^\infty=0$ or not. The `probability' (\ref{IR-RESULT}) matches the classically expected `number of photons' (\ref{IR-CLASSICAL}) with $p'=\pi(p;\infty)$, also as for bremsstrahlung. The removal of this divergence is discussed in the conclusions. To understand the physical differences and similarities between Compton scattering, nonlinear Compton and bremsstrahlung, it is helpful to consider the perturbative limit of our results.

\subsection{Perturbative expansion} \label{SEC:PERT}

\begin{figure}[t!]
	\includegraphics[width=0.4\textwidth]{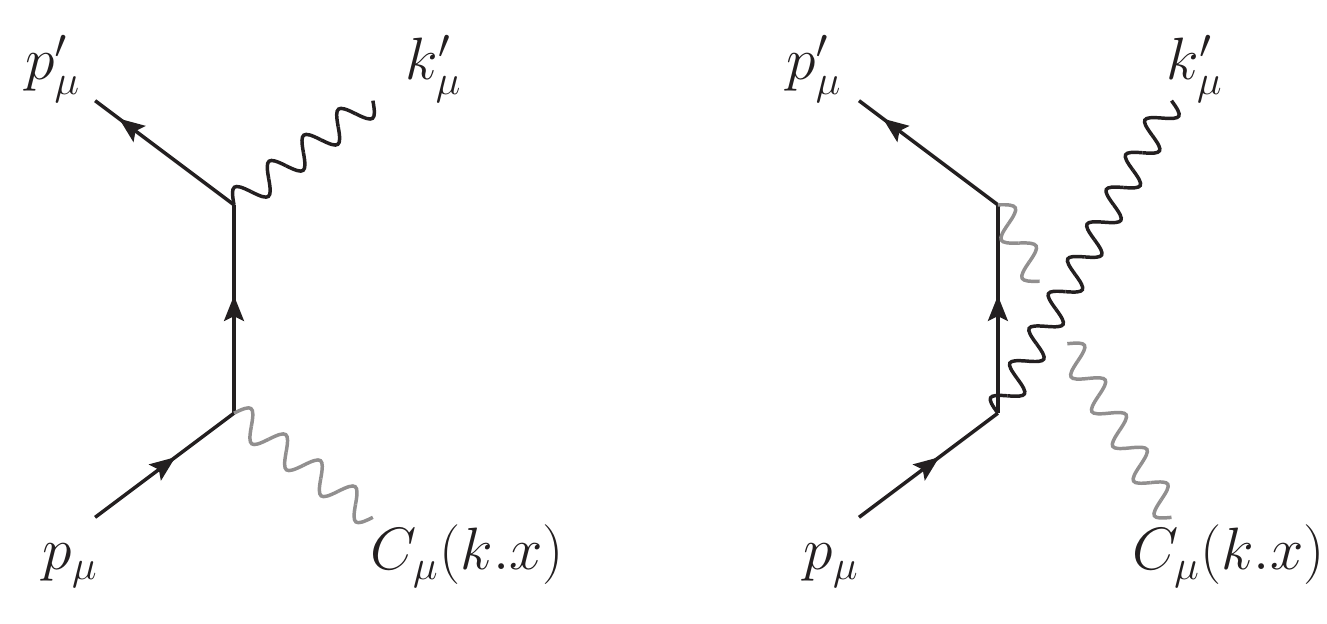}
	\caption{\label{NLC-SFI} Nonlinear Compton scattering at tree level, to lowest order in the background field.}
\end{figure}

%
We assume the incoming electron is at rest in order to give the clearest results. To lowest order in the background, the probability (\ref{THE-PROB}) then becomes
\be\label{NLCP}
 \begin{split}
	\mathbb{P}_\text{pert} &= a_0^2\int\limits_0^\infty\!\frac{\ud s}{2\pi}\ \frac{|\widetilde{f'_j}(s)|^2}{s} \times \\
	&\times \frac{\alpha}{2}\int\limits_{-1}^{1}\!\ud(\cos\theta)\ \bigg(\frac{\omega'_s}{s\omega}\bigg)^2  \bigg[ \frac{\omega'_s}{s\omega} + \frac{s\omega}{\omega'_s}- \sin^2\theta \bigg] \;.
\end{split}
\ee
This is a sum over ordinary Klein-Nishima probabilities for Compton scattering of incoming photons of all frequencies $s\omega$ (second line), modulated by the strength of the background field (first line). The corresponding Feynman diagrams are shown in Fig.~\ref{NLC-SFI}.  The photon frequencies which can be produced by each $s$ are
\be\label{OMEGA-NLC}
\begin{split}
	\frac{1}{\omega'_s} &:= \frac{1}{s\omega} + \frac{1}{m}(1-\cos\theta) \\
	\iff \omega'_s &= \frac{s\omega}{1+\frac{s\omega}{m}(1-\cos\theta)}\;.
\end{split}
\ee
The essential difference between Compton and nonlinear Compton is the range of produced photon frequencies,
\be\label{FULL-RANGE}
	\frac{s\omega}{1+2\tfrac{s\omega}{m}} < \omega' < s\omega \;.
\ee
In Compton scattering one obtains only the second line of (\ref{NLCP}) with $s=1$. The fixed and nonzero incoming photon frequency $\omega$ then acts as an `IR cutoff', since it forbids, via momentum conservation, the outgoing photon from having zero frequency: one obtains (\ref{FULL-RANGE}) with $s=1$. In nonlinear Compton, though, the background field contains a range of frequencies, and each can lead to photon production in the range (\ref{FULL-RANGE}). Even though the range for each $s$ is bounded, $s$ is continuous with $s\geq 0$ and so the emitted photons can in principle be {\it arbitrarily soft}; this is just as in bremsstrahlung, but {\it not} as in Compton scattering.

Whether or not the point $s=0$ can contribute depends, of course, on the low-frequency composition of the beam. For any compactly supported field, i.e.\ a pulse, we can expand $\tdf(s) = \tdf(0) + \mathcal{O}(s) $ for small $s$. From (\ref{OMEGA-NLC}) we have $\omega'_s/s\omega =  1 + \mathcal{O}(s)$, and we find that the soft contribution to the probability (\ref{NLCP}) is
\be\label{NLCP2}
	\mathbb{P}_\text{pert} = \frac{8\alpha a_0^2}{6} \int\limits_0\!\frac{\ud s}{2\pi}\ \frac{|\widetilde{f'_j}(0)|^2}{s}  + \mathcal{O}(s) \;.
\ee
We again obtain the result that the probability is log divergent at $s=0$, corresponding to the emission of a zero-frequency photon, when $\tdf(0)\not=0$. Hence, we confirm that the IR divergence can be attributed to the Fourier zero mode (the zero frequency mode) of the background field strength, this mode permits the production of a zero-frequency photon in a kind of `forward scattering'. This coincides exactly with the ability of the field to accelerate the particle following (\ref{FINT}).

We now turn again to crossed fields.
%
\subsection{Example: crossed fields} \label{SEC:CROSSED}
%
Homogeneous fields are often rather special cases when it comes to radiation, see \cite{Coleman:1961zz} and \cite[\S 37]{SchwingerBook}, and crossed fields are no exception, as we now show.

Clearly, crossed fields can accelerate particles and one therefore expects a soft IR divergence, recall also (\ref{SOFT-J2}). Despite this, the literature results state that the nonlinear Compton probability in crossed fields is IR finite \cite{Nikishov:1963,Nikishov:1964a,Elkina:2010up}.    This probability is given in the appendix. The conclusion from there is that the literature results are indeed IR finite but {\it cannot} be obtained from the limiting case of a homogeneous field of large duration, which does contain an IR divergence. We do not need the details: we can show using only classical arguments that the reason for the difference between the literature results and our own is a different choice of boundary conditions. The literature results describe the crossed field, from the outset, as persisting for all time, and with a gauge potential $\A_\mu(\phi) = a^1_\mu \phi\equiv a_\mu\phi$  \cite{Nikishov:1964a}. The Volkov solutions for this potential, used in the quantum calculation of \cite{Nikishov:1964a}, carry the kinematic momentum
\be\label{TRASH}
	\pi_\mu^c(\phi) = p_\mu - e a_\mu \phi +  \frac{2ea.p\,\phi - e^2a.a\,\phi^2}{2k.p} k_\mu\;,
\ee
and the path $x^c$ is the indefinite integral of $\pi^c_\mu$, where `$c$' stands for crossed.  The (`unregulated') classical current corresponding to the $S$-matrix element of \cite{Nikishov:1963,Nikishov:1964a} is
\be\label{CROSSED-CURRENT}
	j^c_\mu = \frac{e}{k.p}\int\limits_{-\infty}^\infty\!\ud\varphi\ \pi^c_\mu(\varphi)\ e^{ik'.x^c(\varphi)} \;.
\ee
It follows that $p_\mu$ is the kinematic momentum at $\phi=0$; for a particle to have finite momentum after spending an infinite time in the crossed field means that the particle must have {\it started} with infinite momentum. This is clear from (\ref{TRASH}). There are two possible interpretations.

First, one can protest that the field should be considered to turn on at finite time, say $\phi=0$ so that $p_\mu$ is the incoming momentum. In this case, (\ref{CROSSED-CURRENT}) and the corresponding quantum results in the literature contain unphysical contributions from before the particle entered the pulse. Removing them reveals the soft divergence.

Second, one can take (\ref{TRASH}) and (\ref{CROSSED-CURRENT}) at face value. In principle there is no need to regulate the current since the particle never enters or leaves the crossed field. If we did regulate as above, the difference between the two expressions would be a boundary term which would yield a divergence as $\omega'\to 0$. We have seen that understanding the boundary terms (\ref{SOFT-J}) is key to understanding the IR, so let us calculate them. The particle described by (\ref{TRASH}) has infinite kinetic momentum in both the asymptotic past and future, with the leading term being
\be
	\pi^c_\mu(\phi) = -e^2\frac{a.a\,\phi^2}{2k.p}k_\mu + \ldots
\ee
For low frequencies, the boundary term which would cause a soft divergence is therefore
\be
	\lim_{\phi\to\infty} \bigg(\frac{\pi^c_\mu(\phi)}{k'.\pi^c(\phi)} - \frac{\pi^c_\mu(-\phi)}{k'.\pi^c(-\phi)} \bigg) = \frac{k_\mu}{k'.k} - \frac{k_\mu}{k'.k} = 0 \;.
\ee
This means that, from the point of the view of the radiation formulae, the momenta in the asymptotic past and future are not only lightlike but {\it equal}:  since the particle is decelerated from and reaccelerated to the speed of light, there is effectively no net acceleration, and hence no IR divergence\footnote{$S$-matrix elements in crossed fields can be obtained from the low-frequency limit of those in monochromatic waves \cite{Nikishov:1963}. Hence, crossed fields might be considered as `locally constant' approximations for low frequency lasers. Monochromatic waves can themselves be obtained as the limit of $N$-whole-cycle wavetrains when $N\to\infty$ \cite{68016, Heinzl:2010vg}: these do not accelerate, which gives a rather more convoluted explanation for why crossed fields yield IR finite results.}. 

In summary, the literature results for nonlinear Compton scattering in crossed fields are indeed IR finite, but {\it only} on the assumptions that 1) the electron begins with an infinite momentum in the past, and 2) it is decelerated from and then reaccelerated to the speed of light over an infinite time. Dropping these assumptions amounts to allowing the particle to enter and exit the pulse at finite times, and the IR divergence reappears. We leave it to the reader to decide which scenario is more physical.
%
%
\section{Discussion and conclusions}  \label{SEC:CONCS}
%
%
\begin{figure}[b!]
\centering\includegraphics[width=0.4\columnwidth]{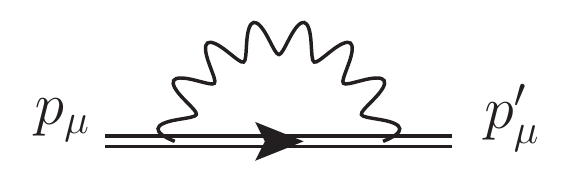}
\caption{\label{FIG:LOOP} One-loop contribution to scattering in a background field without emission. One sees by expanding in powers of the background field that the diagram combines both self-energy and vertex corrections.}
\end{figure}
The IR problem in QED is first encountered in perturbation theory, at tree level, in the bremsstrahlung amplitude for a particle decelerated by an external Coulomb potential. Replacing this potential with a plane wave, we have seen that the same IR divergence is found in nonlinear Compton scattering when the plane wave can give a net acceleration to a particle passing through it (recall that bremsstrahlung  is `breaking radiation').

Our results hold independently of the strength of the background field, and for arbitrary pulse shapes. The only case for which we have allowed a nonvanishing asymptotic field strength is crossed fields. The literature results for nonlinear Compton scattering in crossed fields are, surprisingly, IR finite. We have shown that this results from assuming somewhat unconvincing, unphysical boundary conditions for the scattered particles.

In order to obtain finite and and measurable results for nonlinear Compton scattering, soft emission and higher loop effects must be accounted for.  Tree-level results for the production of one hard photon and an {\it arbitrary} number of soft photons have been calculated and follow the expected IR structure of QED \cite{Linda}: thus, the cancellation of IR divergences to all orders is expected to go through as normal. (See \cite{Glauber:1951zz} for an example of how such structures arise naturally in exactly solvable systems, and also \cite{DiPiazza:2010mv}.) To lowest order in perturbation theory, this cancellation requires adding the calculated probability (\ref{THE-PROB}) of nonlinear Compton scattering to that of scattering without emission, to one loop. The required diagram is shown in Fig.~\ref{FIG:LOOP}. The loop has never been calculated for general plane waves (for crossed fields, see \cite{Ritus:1972ky} and references in \cite{Meuren:2011hv}, for monochromatic fields see \cite{Hartin}), and it will be interesting to investigate both its UV and IR structures when the background is treated nonperturbatively.  This will be discussed in a sequel paper.

Let us finally address the impact of our results on nonlinear Compton scattering in whole-cycle pulses. We saw in the introduction, recall (\ref{LIMIT}) and (\ref{why}), that the total probability of photon production exceeds unity (since the probability of scattering without emission is already unity.)  In Sect.~\ref{SEC:PERT} we saw that even whole-cycle pulses, which give IR finite results, can produce photons with arbitrarily low frequencies when the Fourier spectrum of the pulse extends down to zero frequency. It follows that no detector of finite resolution can distinguish between sufficiently soft emission via nonlinear Compton and scattering without emission (just as soft bremsstrahlung cannot be distinguished from scattering without emission). Experimentally indistinguishable processes must therefore still be accounted for in order to yield measurable probabilities and cross sections for nonlinear Compton experiments, even when the IR divergence is absent.

\acknowledgements
%


V.D.\ acknowledges the support of CNCSIS-of UEFISCSU, project number 488 PNII-IDEI 1909/2008. A.~I. is supported by the Swedish Research Council, contract 2011-4221.

The authors thank Madalina Boca, Viorica Florescu, Chris Harvey, Martin Lavelle, Linda Linsefors, Antonino Di Piazza and Greger Torgrimsson for advice and useful discussions.  Feynman diagrams created using JaxoDraw \cite{Binosi:2003yf,Binosi:2008ig}.


\appendix

\section{Carrier phase}
\begin{figure}[t!]
\includegraphics[width=0.8\columnwidth]{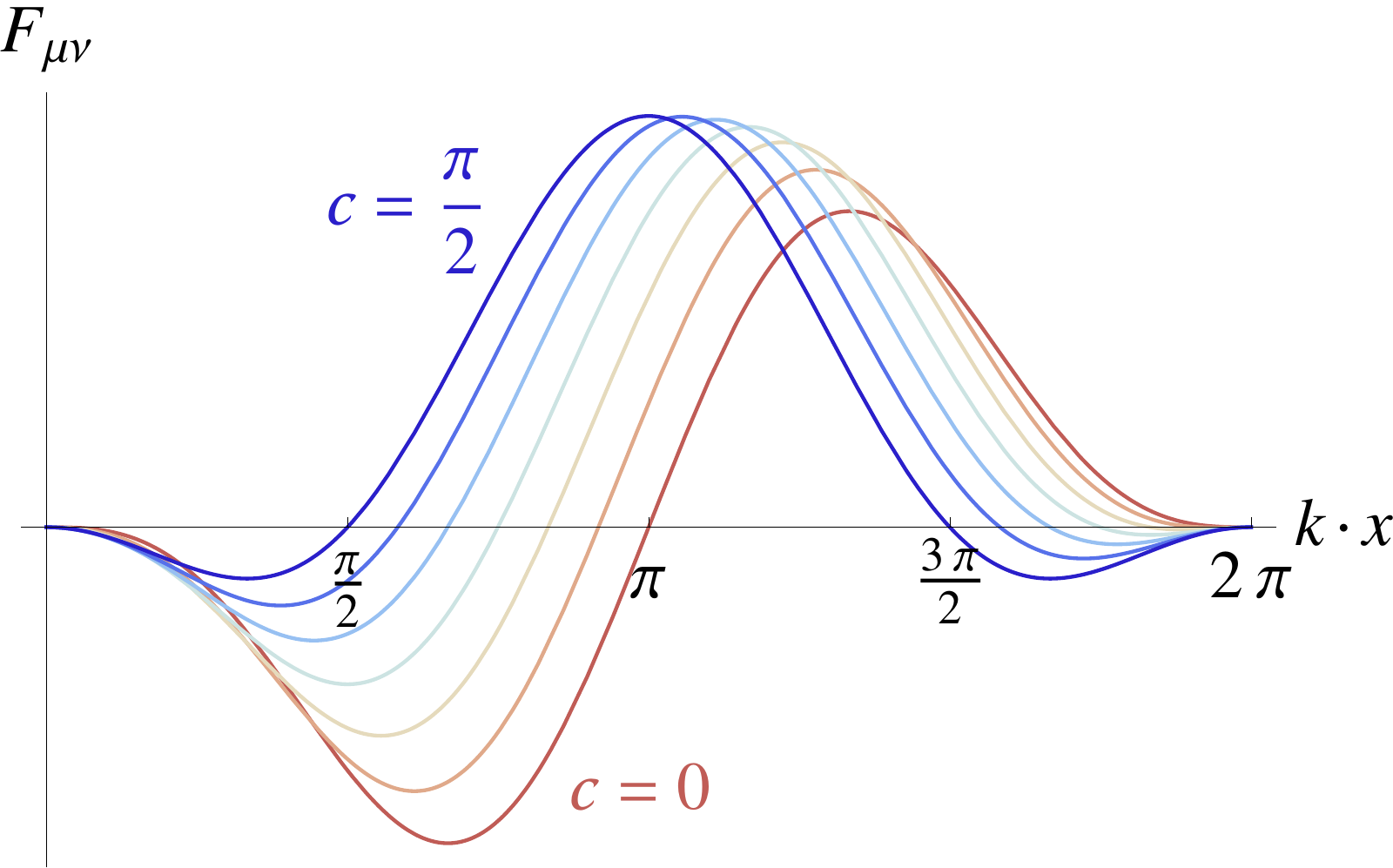}
\includegraphics[width=0.8\columnwidth]{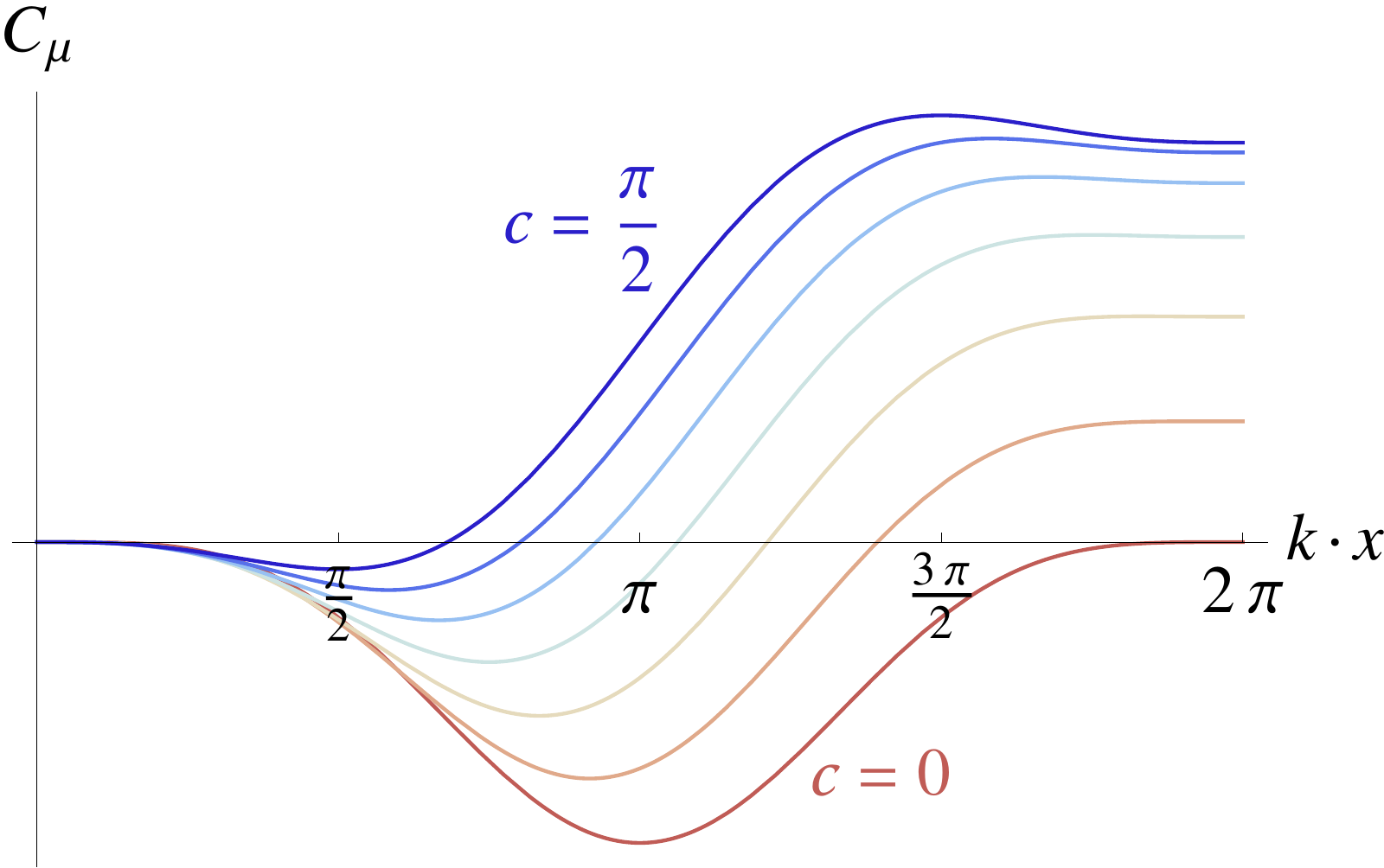}
\caption{\label{FLOWS} The profiles for field strength, $f'(\phi)$ (upper panel), and potential, $f(\phi)$ (lower panel), of a single-cycle pulse with carrier phase $c\in\{0,\pi/2\}$ [red to blue/bottom to top], see equation (\ref{carrier-pulse}). For all $c\not=0$, the potential is non-vanishing when the pulse turns off, implying a net acceleration. }
\end{figure}
\begin{figure}[bt!]
\includegraphics[width=\columnwidth]{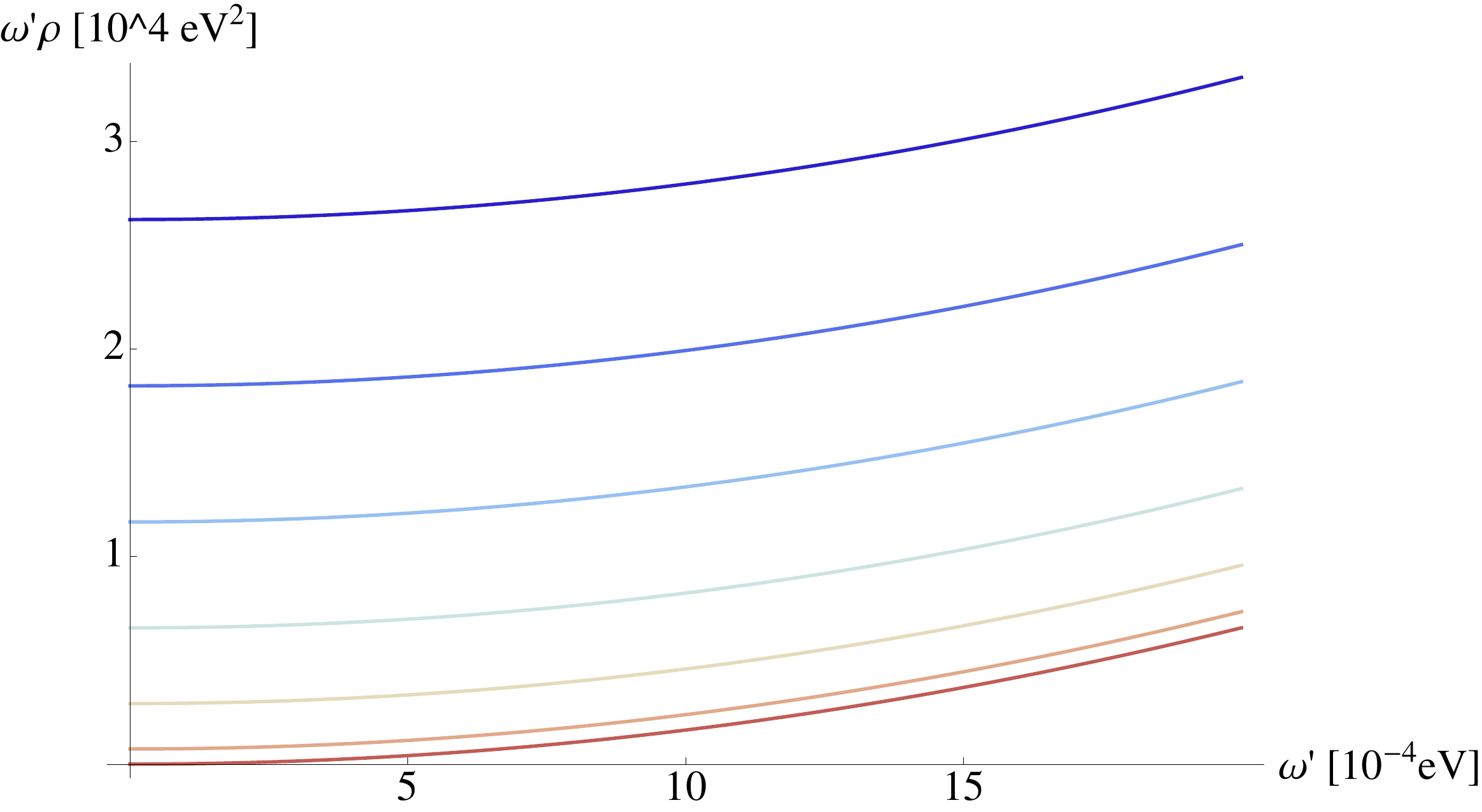}
\caption{\label{CARRIER-PLOTS} The IR behaviour of the energy density $\omega'\rho(\omega')$ in the pulse (\ref{carrier-pulse}), at fixed emission angles. Electron initially at rest, $a_0=1$, backscattered radiation. The carrier phase $c$ is in the range $6 \times 10^{-3}$ [top/blue] to 0 [bottom/red]. The energy density is zero at $\omega'= 0$ only for zero carrier phase.}
\end{figure}

The following simple example provides the quantitative results behind Fig.~\ref{EXEMPEL}. Consider a short pulse with field strength profile
\begin{equation}\label{carrier-pulse}
	f'(\phi) = -n(c) \sin(\tfrac{\phi}{2})^2 \sin (\phi+c) \;,
\end{equation}
for $0\leq \phi\leq 2\pi$ and zero otherwise. The parameter $c$ can be thought of as a `carrier phase' \cite{Mackenroth:2010jk} (see also \cite{Dumlu:2010vv}), and $n(c)$ is a normalisation which ensures the pulse energy is independent of the carrier phase. The field strength and potential $\A_\mu$ are plotted in Fig.~\ref{FLOWS}.  For $c=0$ the field describes a compressed sinusoidal cycle. As $c\to \pi/2$ the pulse acquires a typical `half cycle' shape, see \cite{HALF} for experimental applications of such pulses. For non-zero $c$,  $\A_\mu$ becomes constant and non-zero when the fields turn off, as is also shown. In Fig.~\ref{CARRIER-PLOTS} we display the classical energy density in these pulses as a function of $\omega'$ for small $\omega'$. The former goes to a nonzero constant for all $c\not=0$, and to zero when $c=0$. From (\ref{IR-CLASSICAL}), it is therefore only when $c=0$ (no net acceleration) that the number of photons $N_\gamma$ is finite.

\section{Choice of basis}
Our LSZ reduction formulae allow gauge potentials to be non-zero, but constant, at infinity, and this gives us Volkov solutions labelled by physical momentum and spin. We will show in this section that the same probabilities are obtained if one uses {\it incoming} wavefunctions in (\ref{volkovs}) for both incoming and outgoing electrons, {\it provided} the final electron is integrated out. In other words, we show that the choice of wavefunction is just a choice of basis. Expressing $S$-matrix elements in terms of only incoming variables typically yields more compact expressions, even though it obscures the physics.

\begin{figure}[t!]
\centering\includegraphics[width=0.45\columnwidth]{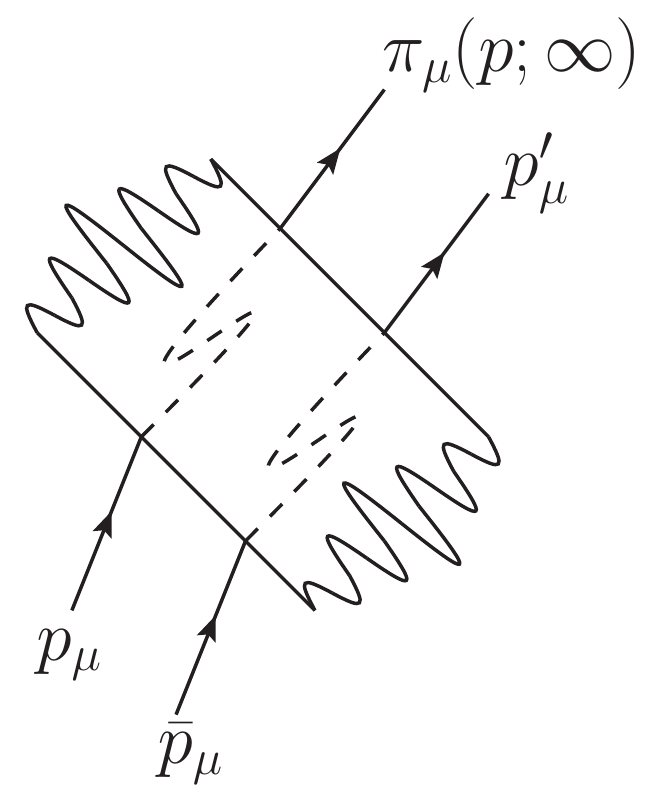}
\caption{\label{FUTUREPAST} In the absence of QED interactions, a particle entering the wave with momentum $p_\mu$ leaves with momentum $\pi_\mu(p;\infty)$. A particle which therefore passes through the wave and {\it leaves} with momentum $p'_\mu$ had a momentum $\bar{p}$ when it entered, where $p'_\mu = \pi_\mu(\bar{p},\infty)$.}
\end{figure}

We will establish the equivalence at the level of the probability (\ref{THE-PROB}), rather than the amplitude level. This means that phase factors generated by our transformations can be neglected (since such phases are always finite in our approach). We begin by introducing a new variable $\bar{p}$ which obeys %
\be
	\pi_\mu(\bar{p};\infty) = p'_\mu \;.
\ee
In other words, $\bar{p}$ is, in the absence of emission, the momentum a particle had {\it before it entered the wave}, if it leaves with momentum $p'$. Explicitly, $\bar{p}$ is
\be\label{PBAR}
	\bar{p}_\mu = {p'}_\mu + e {\A}^\infty_\mu+ \frac{-2 e {\A}^\infty\!\cdot\!p' - e^2 {\A}^\infty.\A^\infty }{2 k\!\cdot\! p'} \, k_\mu \;,
\ee
which is obtained from (\ref{UMU}) by sending $p\to p'$ and $e\to -e$; this is reminiscent of crossing symmetry and amounts to evolving the particle with $p'$ `back in time', through the field, to identify the momentum $\bar{p}$ it began with: see Fig.~\ref{FUTUREPAST}. The expression (\ref{PBAR}) can be derived from the momentum conservation law for scattering without emission (\ref{SFI-ELASTIC0}), by squaring up with $\A^\infty$ on the left hand side, so that the support depends on outgoing $p'$ rather than incoming $p$. Momentum conservation then becomes the requirement that $\bar{p} = p$.
We now turn to nonlinear Compton. Starting with (\ref{THE-PROB}), three transformations are needed, as follows. {\it i) Momentum:}  Change variables,
\be
	s \to s + \frac{2e\A^\infty.p'-e^2\A^\infty.\A^\infty}{2k.p'}k_\LCp.
\ee
This trades $p'$ for $\bar{p}$ in $\Phi$ and the delta-function, and removes the {\it explicit} dependence on $\A^\infty$ therein. {\it ii)Spin:} We consider only probabilities summed over final spins. This spin sum gives
\be\begin{split}
	&\bigg(1+\frac{e\slashed{k}\slashed{\delta\A}}{2k.p'}\bigg)(\slashed{p'}+m)\bigg(1+\frac{e\slashed{\delta\A}\slashed{k}}{2k.p'}\bigg) = \\
	&= \bigg(1+\frac{e\slashed{k}\slashed{\A}}{2k.\bar p}\bigg)(\slashed{\bar p}+m)\bigg(1+\frac{e\slashed{\A}\slashed{k}}{2k.\bar p}\bigg) \;,
\end{split}
\ee
which is the sum one obtains from incoming Volkov wavefunctions with momentum $\bar{p}$. {\it iii) Final states:}
${\bar p}_\mu$ and $p'_\mu$ are two on-shell momenta related by the Lorentz transformation
\be\label{LT}
	 {\Lambda^\mu}_\nu = \exp \bigg[ \frac{e}{k.p'}({\A^\infty k - k \A^\infty)^\mu}_\nu \bigg] \;,
\ee
and, despite the momentum dependence of this transformation, the measure over final states is invariant under ${\Lambda^\mu}_\nu$ \cite{Brown:1969nb}, so that
\be
	\int\!\frac{\ud^3{\bar p}}{2{\bar p}_0} = 	\int\!\frac{\ud^3 p'}{2p'_0} \;.
\ee
Performing these manipulations, (\ref{THE-PROB}) reduces to the result one would obtain obtain by using the same Volkov solutions to describe both incoming and outgoing particles, as has appeared in the literature to date. In summary, the `LSZ approach' tells us to use different bases for incoming and outgoing states. These bases are labelled by physical momenta, etc. There is nothing to stop us, though, from expanding out-states in a basis of in-states: this is implicitly done when one uses the same Volkov wavefunctions for both incoming and outgoing particles. (There is no distinction for whole-cycle pulses.) Both approaches yield the same results when the outgoing electron degrees of freedom are integrated out. However, if one is interested in {\it differential} rates or probabilities with respect to the electron momentum, one should change variables from $\bar{p}$ back to the physical $p'_\mu$.

\section{Probabilities and crossed fields}\label{Alt-sect}
Using the above results, the nonlinear Compton probability is most compactly written
\be\label{PROB}
\begin{split}
	\mathbb{P} = \frac{e^2m^2}{k.p} & \int\!\frac{\ud^3 \bar p}{(2\pi)^32{\bar p}_0} \int\!\frac{\ud^3 k'}{(2\pi)^3 2\omega'} \\
	& \int\!\frac{\ud s }{2\pi}\ (2\pi)^4 \delta^4(\bar{p}+k' - p - sk)\mathcal{J} \;.
\end{split}
\ee
Let $x\equiv k.k'/k.p'$. All dependence on the pulse profile is contained in
\be\begin{split}
	\mathcal{J} = &-2|B_0|^2 + a^2\bigg(1 + \frac{x^2}{2(1+x)}\bigg)\times \\
	& \times \big(2|{B}_1|^2 + 2|{B}_2|^2-{B}_0{B}^*_3-{B}_0^*{B}_3\big) \;,
\end{split}
\ee
through four functions $B_\mu$,
\be\label{B-MU}
	B_\mu = \int\ud\phi\, e^{i\Phi}\frac{\ud}{\ud\phi} \bigg(\frac{f_\mu(\phi)}{i\Phi'}\bigg) \;,
\ee
where we define $f_0\equiv 1$, $f_3\equiv f_1^2+f_2^2$ and the phase $\Phi$ is now
\be
	\Phi(x) := s x - \alpha_j\! \int\limits^{x}_{-\infty}\!\ud y\ f_j(y) \;,
\ee
with $j$ summed over $\{1,2,3\}$. The $\alpha$ parameters are constructed from the incoming Volkov solutions,
\be\begin{split}\label{alphas}
	\alpha_j &= e a^\mu_j \!\cdot\!\bigg( \frac{p_\mu}{k.p} - \frac{{\bar p}_\mu}{k.{\bar p}}\bigg) \;, \quad j = 1,2 \;, \\
	\alpha_3 &= -\frac{m^2a_0^2}{2} \frac{k.k'}{k.p \, k.\bar{p}} \;,
\end{split}	
\ee
and the $B_\mu$ obey $sB_0 = \alpha_j B_j$ as a consequence of (\ref{DELTA-GAMMA}).

Our assumption on the behaviour of the electromagnetic fields (that they vanish asymptotically) does not allow us to apply the above results to crossed fields directly. Instead we take the limit of the more physical situation in which a particle enters and leaves a patch of constant field strength. We will compare these results with those in the literature which assume a crossed field from the outset. The potential for a field which is constant for a lightfront time $T$ is given in (\ref{CROSSED-POTENTIAL}). Applying the results of the previous subsection, one finds that the functions (\ref{B-MU}) become (changing variables $\varphi= \phi+T/2$)
\be
	B_\mu = -\int\limits_0^T\ud\varphi\  e^{i\Phi_c}\frac{\ud}{\ud\varphi}\bigg(\frac{b_\mu(\varphi)}{i\Phi'(\varphi)}\bigg) \;,
\ee
with $b_\mu(\varphi) = (1,\varphi,0,\varphi^2)$ and the crossed-field phase
\be
	\Phi_c(\varphi) = \left( s+\tfrac{\alpha_1^2}{4\alpha_3} \right) \varphi - \tfrac{\alpha_3}{3} \left(\varphi+\tfrac{\alpha_1}{2\alpha_3}\right)^3 \;.
\ee
In order to compare this result with that in the literature, we integrate by parts - without dropping the boundary term! - to find
\be\label{CROSSED-A}
	B_\mu = - e^{i\Phi_c}\frac{b_\mu(\varphi)}{i\Phi'(\varphi)}\bigg|_0^T +\int\limits_0^T\ud\varphi\  e^{i\Phi_c} b_\mu(\varphi)\;,
\ee
As $\omega'\to 0$, and using momentum conservation, the boundary term survives, reproducing the IR divergence of (\ref{SOFT-J2}), while the second, `bulk' term vanishes. Now consider the limit as $T\to\infty$. The only $T$-dependence in the bulk term is in the integral limit, so we replace $T\to\infty$ there: this should be compared with the corresponding literature expression for $B_\mu$, which is \cite{Nikishov:1963,Nikishov:1964a}
\be\label{CROSSED-NR}
	B_\mu \overset{!}{=} \int\limits_{-\infty}^\infty\!\ud \varphi\ e^{i\Phi_c} b_\mu(\varphi)  \;.
\ee
(To convert from the conventions of \cite{Nikishov:1963,Nikishov:1964a} to ours, use $p'_\text{N.R.} \to \bar{p}$, $\alpha_\text{N.R.}\to -\alpha_1$, $s_\text{N.R.}\to -s$ and $\beta_\text{N.R.} \to 4\alpha_3$.) The results (\ref{CROSSED-A}) and (\ref{CROSSED-NR}) are not equivalent, even as $T\to\infty$. The former contains a boundary term giving an IR divergence and has `semi-infinite' integration limits, both of which are a consequence of the particle being allowed to enter and leave the background.  The literature result (\ref{CROSSED-NR}) assumes a constant field from the outset, which the particles never enter or leave. This is consistent, but it shows that (\ref{CROSSED-NR}) cannot be obtained as the large-duration limit of (\ref{CROSSED-A}). Further discussion may be found in Sect.~\ref{SEC:CROSSED}.

\end{document}